\documentclass[aps,nofootinbib,showpacs,showkeys,preprint
,tightenlines ] {revtex4}

\usepackage{epsf,epsfig,subfigure,graphicx,amsmath,amssymb}
\usepackage{color}

\renewcommand{\a}{{\sf a}}
\newcommand{\tr}{{\rm tr}}
\renewcommand{\v}{\check}
\newcommand{\Z}{{\mathbb Z}}

\begin{document}

 \begin{flushright}
{\tt KIAS-P11002
\\PNUTP-11-A02}
\end{flushright}

\title{\Large\bf Weak Mixing Angle and Proton Stability \\
in F-theory GUT}

\author{ Kang-Sin Choi$^{(a)}$\footnote{email: kschoi@kias.re.kr}
and Bumseok Kyae$^{(b)}$\footnote{email: bkyae@pusan.ac.kr} }
\affiliation{$^{(a)}$ School of Physics, Korea Institute for Advanced
Study, Seoul 130-722, Korea
\\
$^{(b)}$ Department of Physics, Pusan National University, Busan
609-735, Korea}


\begin{abstract}
It is pointed out that a class of flipped SU(5) models based on
F-theory naturally explains the gauge coupling unification. It is because the group SU(5)$ \times $U(1)$_X$ is embedded in SO(10) and $E_8$. To prohibit
the dimension 4 and 5 proton decay processes,
the structure group should be SU(3)$_\perp$ or
smaller. Extra heavy vector-like pairs of $\{{\bf
5}_{-2},{\bf\overline{5}}_2\}$ except only one pair of Higgs
should be also disallowed, because they could induce the unwanted
dimension 5 proton decays. We construct a simple global F-theory
model considering these points. To maintain ${\rm
sin}^2\theta_W^0=\frac38$ at the GUT scale, the fluxes are
turned-on only on the flavor branes.
\end{abstract}

 \pacs{11.25.Mj, 11.25.Wx, 12.10.Kt}
 \keywords{F-theory, Gauge coupling unification, Weak mixing angle, Proton decay}
 \maketitle


\section{Introduction}

One of the dramatic successes in the Minimal Supersymmetric
Standard Model (MSSM) is the gauge coupling unification. Thanks to
the additional contributions by the superpartners to the
renormalization effects, the three gauge couplings of the MSSM,
$\{g_3, g_2\sqrt{\frac53}g_Y\}$ can be unified quite accurately at
$2\times 10^{16}$ GeV energy scale \cite{MSSMunif}.\footnote{For
the hypercharges of the MSSM superfields, we take the convention
of $Y[q]=\frac16$, $Y[u^c]=-\frac23$, $Y[l]=-\frac12$, etc.
throughout this paper. In our notation, $q$, $u^c$($d^c$), $l$,
and $e^c$($\nu^c$) mean the quark doublet, quark singlet with
$Q_{\rm em}=-\frac23$($+\frac13$), lepton doublet, and lepton
singlet with $Q_{\rm em}=+1$($0$), respectively. For the
superheavy fields carrying the same quantum numbers with the MSSM
fields, mainly the capital letters will be utilized in this
paper.} It seems to imply the presence of a supersymmetric (SUSY)
unified theory at that scale. When discussing the gauge coupling
unification in the MSSM, however, one should notice that such a
unification is possible, since the normalization for $g_Y$ (and
also the normalization for the hypercharges) deduced in SU(5) and
SO(10), i.e. $\sqrt{\frac53}$ ($\sqrt{\frac35}$) is employed. This
normalization predicts that the weak mixing angle, which is
defined as ${\rm sin}^2\theta_W\equiv\frac{g_Y^2}{g_2^2+g_Y^2}$,
should be $\frac38$ at the unification scale.

One of the problems in SUSY grand unified theories (GUTs) is the
doublet/triplet splitting in the Higgs multiplets. Unlike in the
matter sector, the electroweak Higgs in the MSSM, $\{h_u, h_d\}$
can be embedded in proper GUT multiplets [e.g. $\{{\bf
5},{\bf\overline{5}}\}$ in SU(5) and ${\bf 10}$ in SO(10)] with
unwanted SU(3) triplets $\{D,D^c\}$ supplemented. Although they
are contained in a common multiplet, how to make the triplets
superheavy while keeping the doublets massless down to the
electroweak scale are known to be a notorious problem in GUT.

This problem is closely associated also with the proton decay in
SUSY GUTs \cite{Nath}. While the dimension 4 proton decay
processes can be prohibited by introducing the R-parity, the
dimension 5 processes can not be forbidden by it.  This problem
arises often also in the minimal SU(5) and SO(10) in other guises.
Even though one successfully splits the doulet/triplets, unless
the triplet pieces of the Higgs multiplets are decoupled by an
elaborate way, the operators leading to the dimension 5 proton
decay
%
are generated again at tree level.

Flipped SU(5), which is based on the gauge group
SU(5)$\times$U(1)$_X$, provides very nice framework addressing
these problems \cite{flippedSU5,missingpartner}. In flipped SU(5),
the ``missing partner mechanism'' for doublet/triplet splitting
works in a very simple way \cite{missingpartner}. Such split
triplets do not induce the dimension 5 proton decay in flipped
SU(5). Moreover, in flipped SU(5) there is no serious fermion mass
relations constraint by the GUT group structure, which arise often
in many simple GUTs. However, the gauge group of flipped SU(5) is
a semi-simple group. Thus, it can address the gauge coupling
unification, only when it is embedded in a promising UV theory
such as string theory; it could determine the U(1)$_Y$
normalization such that ${\rm sin}^2\theta_W^0=\frac38$ at the GUT
energy scale \cite{stringFlip,KimKyae}.

In this paper, we attempt to construct a flipped SU(5) model based
on F-theory. We will point out that the predicted $\sin^2\theta_W^0$ at the string scale, which is assumed to be
around the GUT scale, is $\frac38$.\footnote{In the strongly
coupled heterotic string theory (or heterotic M-theory), the
fundamental scale becomes coincident with the GUT scale
\cite{heteroticM}. As dual to the heterotic
M-theory, F-theory has the same relation.} Hence the three gauge couplings in the MSSM [or SU(5)
and U(1)$_X$ gauge couplings] are unified at the GUT scale. In
order to obtain the chiral fields in 4 dimensional spacetime (4D)
and to maintain the gauge coupling unification, we will turn on
the universal fluxes only on the flavor branes. We will also
discuss how to forbid dimension 4 and 5 proton decay processes in
the flipped SU(5) model based on F-theory such that the dimension
6 process [$p\to e^+\pi^0$ with $\tau_p\approx 10^{34-35}$yr]
becomes the dominant one.

F-theory is defined by lifting the $SL(2,\Z)$ symmetry of Type IIB string theory to that of geometric torus.
The axion-dilaton field in IIB string is identified to the complex
structure of the torus \cite{Vafa,Reviews}.
Toward a four dimensional $N=1$ SUSY model, we compactify F-theory on Calabi-Yau fourfold, which is elliptically fibred on a three-base $B$.

The varying axion-dilaton field on $B$, which naturally
incorporates non-perturbative effects, makes more light degrees of
freedom than open fundamental strings possible, so that
exceptional group of $E_n$ series emerges. Identifying
$E_3,E_4,E_5$ as SU(3)$\times$SU(2), SU(5) and SO(10),
respectively, we have natural symmetry enhancement patterns $E_3
\times$U(1)$_Y \times$U(1)$_X \subset E_4 \times$U(1)$_X \subset
E_5$ \cite{BHVI,BHVII,DWI}. Thus, F-theory enables us to track how
such unification pattern is realized.

In particular $E_4\times$U(1)$_X$ naturally provides the flipped SU(5) group: not only gauge group but also matter contents and Yukawa couplings nicely
fit \cite{BHVII,flipFmodels,KuMa}. As mentioned above, we will try to
construct an F-theory model to reproduce a field-theoretically
desired flipped SU(5) model, particularly addressing the issues on
the gauge coupling unification and the absence of dimension 4 and
5 proton decay operators \cite{BHVII,ChoiKim}.
Conventional GUTs employ Higgs scalar fields to break GUT groups
to the SM group. The Higgs mechanism in SUSY GUTs could inherit
the gauge coupling unification of the MSSM. In F-theory GUT, there
is another way of GUT breaking using flux. A flux along the
hypercharge direction, however, is known to distort a little bit
the gauge coupling unification \cite{distort,DWII}. To track the origin
of the observed value of ${\rm sin}^2\theta_W$, we will consider
an F-theory GUT as  not $E_4$ but $E_4\times$U(1), whose breaking solely relies on the Higgs mechanism.

This paper is organized as follows. In section II, we will briefly
review flipped SU(5) and discuss dimension 4 and 5 proton decay in
flipped SU(5). In section III, we construct an F-theory model of
flipped SU(5). When constructing a model, we will particularly
focus on how to reflect the gauge coupling unification observed in
the MSSM, and to avoid the dimension 4 and 5 proton decay
processes. In section IV, we will discuss low energy physics
expected from our F-theory model. Section V will be devoted to
conclusions.

\section{Flipped SU(5)}

The gauge group of flipped SU(5) is SU(5)$\times$U(1)$_X$. Unlike
in the conventional SU(5) i.e. Georgi-Glashow's SU(5) [$\equiv$
SU(5)$_{\rm GG}$] \cite{GG}, the hypercharge of the standard model
(SM) is defined as a linear combination of a diagonal SU(5) and
U(1)$_X$ generators:
\begin{eqnarray} \label{hyperY}
Y=\frac15\left(T_5+X\right) ~,
\end{eqnarray}
where $T_5$ [$={\rm
diag.}(\frac13~\frac13~\frac13~\frac{-1}{2}~\frac{-1}{2})$] is a
diagonal generator of SU(5), and $X$ denotes the U(1)$_X$ charge.
For a while, let us neglect the normalizations of $T_5$ and $X$.
Table \ref{tb:contents} lists the field contents of the flipped
SU(5) model, from which one can see how the MSSM superfields are
embedded there.
%
\begin{table}
\begin{tabular}{c|c}
\hline SU(5)$_X$ ~&~ MSSM fields
\\
\hline ${\bf 10}_1$ & $\{d^c, q, \nu^c\}$ \\
${\bf\overline{5}}_{-3}$ & $\{u^c, l\}$ \\
${\bf 1}_5$ & $e^c$ \\ \hline
${\bf 5}_{-2}$ ($\equiv {\bf 5}_{h}$) ~& $\{D, h_d\}$ \\
${\bf\overline{5}}_2$ ($\equiv {\bf\overline{5}}_h$) & $\{D^c, h_u\}$ \\
\hline
\end{tabular}
\caption{ ~Superfields in flipped SU(5). The SU(3) triplets $D$
and $D^c$ are absent in the MSSM, which should be decoupled from
low physics. When flipped SU(5) embedded in SO(10), the $X$
charges in the table should be normalized as
$X\to\frac{1}{\sqrt{40}}X$. } \label{tb:contents}
\end{table}
%

Breaking of SU(5)$\times$U(1)$_X$ to the SM gauge group demands
introduction of the Higgs fields $\{{\bf 10}_H,
{\bf\overline{10}}_H\}$, which carry the quantum numbers of
$\{{\bf 10}_1, {\bf{\overline{10}}}_{-1}\}$, respectively. Since
they contain the SM singlets $\{\nu^c_H, \overline{\nu}^c_H\}$,
their vacuum expectation values (VEVs) in the SM singlet
directions result in spontaneous breaking of flipped SU(5) to the
SM gauge group.

The different definition of the hypercharge results in the
different embedding of the MSSM fields: comparing with SU(5)$_{\rm
GG}$, $d^c$, $\nu^c$ and $h_d$ are replaced by $u^c$, $e^c$, and
$h_u$, respectively. As a result, the prediction from Yukawa
couplings in flipped SU(5) is also different from that of the
conventional SU(5). The superpotential in flipped SU(5) is written
down as follows:
\begin{eqnarray} \label{Yukawa}
W=y^{(d)}_{\;ij}{\bf 10}_i{\bf 10}_j{\bf
5}_{h}+y^{(u,\nu)}_{\;ij}{\bf
10}_i{\bf\overline{5}}_{j}{\bf\overline{5}}_{h}+
y^{(e)}_{\;ij}{\bf 1}_{i}{\bf\overline{5}}_{j}{\bf 5}_{h}+ \mu{\bf
5}_{h}{\bf\overline{5}}_{h}+\frac{y^{(m)}_{\;ij}}{M_P}{\bf\overline{10}}_H{\bf\overline{
10}}_H{\bf 10}_i{\bf 10}_j ~,
\end{eqnarray}
where $i,j$ stand for the family indices. From the first term,
d-type quarks [rather than u-type quarks as in the SU(5)$_{\rm
GG}$] get masses. From the second term, u-type quarks' and Dirac
neutrinos' masses are generated, and they are related as
$M^{(u)}_{ij}=M^{(\nu)}_{ji}$ [rather than
$M^{(d)}_{ij}=M^{(e)}_{ji}$]. However this relation is not much
crucial, because the physical neutrino masses are given by the
Majorana mass terms as well as the Dirac mass terms. The Majorana
masses are induced by the last term of Eq.~(\ref{Yukawa}), when
${\bf\overline{10}}_H$ develop a VEV in the right-handed neutrino
direction. Thus, there is no effective mass relation in flipped
SU(5), and so unrealistic mass relations predicted in other simple
GUT models are absent. The charged leptons achieve the masses from
the third term of Eq.~(\ref{Yukawa}). From now on, we will provide
some comments on flipped SU(5) in order.

\subsection{Weak Mixing Angle and Coupling Unification}

Normalization of U(1) charges seems arbitrary, since rescaling of the charges can be absorbed by the coupling constant. The same can be true for $X$ of bottom-up constructed flipped SU(5). But this is not the case if U(1)$_X$ is embedded in a simple group, since then U(1) coupling becomes not independent. If SU(5)$\times$U(1)$_X$ is embedded in SO(10),
it should be fixed to $\frac{1}{\sqrt{40}}X$. In such a case,
hence, ${\bf 10}_1$, ${\bf 5}_{-3}$, ${\bf 1}_5$, etc. in Table
\ref{tb:contents} should be replaced by ${\bf 10}_{1/\sqrt{40}}$,
${\bf 5}_{-3/\sqrt{40}}$, ${\bf 1}_{5/\sqrt{40}}$, and so forth.
Even in such a case, however, we will drop the normalization
factor in the subscripts, just tacitly assuming it for simplicity
in notations.
Indeed, the U(1)$_X$ charge normalization by $\frac{1}{\sqrt{40}}$
yields ${\rm sin}^2\theta_W^0=\frac38$, unifying the SU(5) and
U(1)$_X$ gauge couplings at the GUT scale (see e.g. appendix of
Ref. \cite{KimKyae}).

As mentioned in Introduction, there are many difficult problems
such as the doublet/triplet splitting problem of the Higgs sector
in ordinary 4 dimensional SUSY GUTs. Hence, it would be desirable
to construct a flipped SU(5) model in the framework of string
theory such that the normalization of the U(1)$_X$ charges is
given by $\frac{1}{\sqrt{40}}$ \cite{stringFlip,KimKyae}. In that
case, the flipped SU(5) gauge group is embedded in a much larger
group, but it is broken to SU(5)$\times$U(1)$_X$ not by a
spontaneous breaking mechanism but by a way associated with a
compactification mechanism of the extra space dimensions. Such en
explicit construction of flipped SU(5) from string theory with
realizing the desired normalization of U(1)$_X$ could easily avoid
the problems appearing in SUSY GUTs.

\subsection{Missing Partner Mechanism}

Flipped SU(5) can be broken to the SM gauge group by the tensor
Higgs ${\bf 10}_H$ and ${\bf\overline{10}}_H$ carrying the $X$
charges $+1$ and $-1$, respectively. In terms of the SM quantum numbers, the
tensor Higgses ${\bf 10}_H$ and ${\bf\overline{10}}_H$ split to
$\{d^c_H, q_H, \nu^c_H\}$ and $\{d_H, \overline{q}_H, \nu_H\}$,
respectively. When ${\bf 10}_H$ and ${\bf\overline{10}}_H$ develop
VEVs along the $\nu^c_H$ and $\nu_H$ directions, $q_H$ and
$\overline{q}_H$ are absorbed by the heavy gauge sector, but
$d^c_H$ and $d_H$ contained in ${\bf 10}_H$ and
${\bf\overline{10}}_H$ potentially remain as pseudo Goldstone
modes. Somehow they should be made superheavy to protect the gauge
coupling unification.

$\{D, D^c\}$ modes included in $\{{\bf 5}_{h},
{\bf\overline{5}}_{h}\}$ should be also removed from the low
energy field spectrum, while the doublets in $\{{\bf 5}_{h},
{\bf\overline{5}}_{h}\}$ should survive down to low energies
because they are nothing but the electroweak Higgs in the MSSM.
This is the doublet/triplet splitting problem in flipped SU(5).
However, the unwanted $\{d^c_H, d_H\}$ from $\{{\bf 10}_H,
{\bf\overline{10}}_H\}$ and $\{D, D^c\}$ from $\{{\bf 5}_{h},
{\bf\overline{5}}_{h}\}$ turn out to be superheavy by pairing with
each other. It is a merit of flipped SU(5). Consider the following
superpotential,
\begin{eqnarray} \label{D/Tsplit}
W\supset {\bf 10}_H{\bf 10}_H{\bf 5}_h
+{\bf\overline{10}}_H{\bf\overline{
10}}_H{\bf\overline{5}}_h=\langle\nu_H^c\rangle
d^c_HD+\langle\overline{\nu}_H^c\rangle d_HD^c ~,
\end{eqnarray}
which is allowed in flipped SU(5). As seen in
Eq.~(\ref{D/Tsplit}), all the unwanted modes discussed above
become superheavy by obtaining the Dirac masses proportional to
$\langle{\bf 10}_H\rangle$ and
$\langle{\bf\overline{10}}_H\rangle$. However, one should note
that this mechanism works for only one pair of vector-like Higgs
fields. If there are more heavy Higgs-like fields $\{{\bf 5}_{G},
{\bf\overline{5}}_{G}\}$, the triplet modes included there can not
get masses through this mechanism; introducing another pairs
$\{{\bf 10}_H^\prime,{\bf\overline{10}}_H^\prime\}$ for removing
such triplets would leave unwanted pseudo Goldstones
$\{q_H^\prime,\overline{q}_H^\prime\}$ contained in $\{{\bf
10}_H^\prime,{\bf\overline{10}}_H^\prime\}$, which can not eaten
by the gauge sector.

\subsection{Proton Stability}

In the MSSM the baryon and lepton numbers are conserved by
R-parity at the renormalizable level. (It might be an ad hoc
introduction for the baryon and lepton number conservation, and
dark matter.) Even R-parity, however, can not prohibit the
dimension 5 proton decay processes. In flipped SU(5), the R-parity
violating terms in the MSSM do not arise from the renormalizable
superpotential at all, because they are forbidden by U(1)$_X$
[unlike in SU(5)$_{\rm GG}$]. However, such R-parity violating
terms as well as the terms leading to dimension 5 proton decay can
appear from the non-renormalizable superpotential:
\begin{eqnarray}
&&\frac{1}{M_P}{\bf 10}_H{\bf 10}_i{\bf 10}_j{\bf\overline{5}}_{k}
\rightarrow \frac{\langle\nu^c_{H}\rangle}{M_P}\left(q_id^c_jl_k +
d^c_id^c_ju^c_k\right) ~, ~~~~ \frac{1}{M_P}{\bf
10}_H{\bf\overline{5}}_{i}{\bf\overline{5}}_{j}{\bf 1}_{k}
\rightarrow \frac{\langle\nu^c_{H}\rangle}{M_P} l_il_je^c_k ~,
\quad\quad\label{Rp-viol}\\
&&\frac{1}{M_P}{\bf 10}_i{\bf 10}_j{\bf 10}_k{\bf\overline{5}}_{l}
\rightarrow \frac{1}{M_P}q_iq_jq_kl_l ~,
\quad\quad\quad\quad\quad\quad ~~ \frac{1}{M_P}{\bf
10}_i{\bf\overline{5}}_{j}{\bf\overline{5}}_{k}{\bf 1}_{l}
\rightarrow \frac{1}{M_P} d^c_iu^c_ju^c_ke^c_l ~, \quad\quad
\label{D5pdecay}
\end{eqnarray}
where $i,j,k,l$ indicate again the family indices. These terms in
the superpotential should be forbidden somehow for the baryon and
lepton number conservations. Then, proton decay would be dominated
by dimension 6 operators, which are still safe for the proton
longevity. But it it not the end of the discussion.

Let us suppose that there is an extra vector-like pair of $\{{\bf 5}_{G},
{\bf\overline{5}}_{G}\}$, which carries the same quantum numbers
with the electroaweak Higgs pair $\{{\bf 5}_{h},
{\bf\overline{5}}_{h}\}$. Then the allowed superpotential is as
follows:
\begin{eqnarray} \label{unwanted}
W_{\rm unwanted}={\bf 10}_i{\bf 10}_j{\bf 5}_{G}+{\bf
10}_k{\bf\overline{5}}_{l}{\bf\overline{5}}_{G}+ {\bf
1}_{m}{\bf\overline{5}}_{n}{\bf 5}_{G}+ M_G{\bf
5}_{G}{\bf\overline{5}}_{G} ~,
\end{eqnarray}
where $M_G$ is supposed to be a GUT or Planck scale mass
parameter. Hence, the extra pair $\{{\bf 5}_{G},
{\bf\overline{5}}_{G}\}$ achieves a superheavy Dirac Mass $M_G$.
In terms of the SM, ${\bf 5}_{G}$ and ${\bf\overline{5}}_{G}$
split into $\{D_G, L_G\}$ and $\{D^c_G, L^c_G\}$, respectively.
The first three terms of Eq.~(\ref{unwanted}) are presented as
\begin{eqnarray}
\left(d^c_{\{i}\nu^c_{j\}}+q_{\{i}q_{j\}}+e^c_mu^c_n\right)D_G
+\left(d^c_{\{i}q_{j\}}
+e^c_ml_n\right)L_G+(d^c_ku^c_l+q_kl_l)D^c_G
+(q_ku^c_l+\nu^c_kl_l)L^c_G ~, ~~
\end{eqnarray}
Note that after integrating out the heavy $\{D_{G}, D^c_G\}$ modes
included in $\{{\bf 5}_{G}, {\bf\overline{5}}_{G}\}$, the unwanted
terms of Eq.~(\ref{D5pdecay}) are generated again. They are
suppressed by $1/M_G$ (rather than $1/M_P$). Thus, the extra pair
of $\{{\bf 5}_{G}, {\bf\overline{5}}_{G}\}$ are also dangerous for
proton stability, even if they are superheavy.
In the case of the SM gauge symmetry, this problem could arise
also, if there are extra vector-like pairs of heavy SU(3)
triplets.

On the other hand, $\{D,D^c\}$ included in the Higgs multiplets
$\{{\bf 5}_{h}, {\bf\overline{5}}_{h}\}$ become superheavy by
pairing with ${\{d^c_H, d_H\}}$ contained in $\{{\bf 10}_{H},
{\bf\overline{10}}_{H}\}$ as discussed in subsection B, and the
mass parameter corresponding to $M_G$ of Eq.~(\ref{unwanted}),
namely, ``$\mu$'' in Eq.~(\ref{Yukawa}) is just of the electroweak
scale. Accordingly, the terms induced by $\{D,D^c\}$ are
suppressed by $\mu/\langle{\bf 10}_H\rangle^2$ rather than
$1/M_P$, which are extremely small.

\section{Construction from F-theory}

Before constructing a model from F-theory, let us discuss first
some results inferred by considering only the gauge invariance and
the notion of monodromy. The low energy theory would be eventually
embedded in $E_8$: all the SM matter originate from the branching
of its gaugino. Namely, under $E_8\to$SU(5)$\times$
U(1)$_X\times$SU(4)$_\perp$, the adjoint branches as
\begin{equation} \label{branching} \begin{split}
{\bf 248} & \to {\bf
(24,1)}_{0}+{\bf (1,15)}_{0}+{\bf (1,1)}_{0}
\\& +\left[{\bf (1,4)}_{5}+{\bf (5,\overline{6})}_{-2}+{\bf
(\overline{5},4)}_{-3}+{\bf (10,4)}_{1}+{\bf
(10,1)}_{-4}+ {\rm c.c. }\right].
  \end{split}
\end{equation}
Focusing on SU(5)$\times$U(1)$_X$ quantum numbers, we see it
reproduces the desired matter contents in the minimal way. The
only unwanted one is the only SU(4)$_\perp$ singlet ${\bf
(10,1)}_{-4}$. We can easily remove it from low energy field
spectrum just by manipulating $G$-flux. We will discuss it again
later.

We achieve the desired symmetry breaking by embedding a background
gauge bundle of the structure group SU(4)$_\perp\times$U(1)$_X$.
The unbroken group is the commutant group in $E_8$, i.e. the
flipped SU(5) group, SU(5)$\times$U(1)$_X$. Note that U(1)$_X$ can
be unbroken, because it commutes with itself. The issue concerning
its anomaly will be discussed later. The important properties of
the gauge bundle are the followings. First, the zero mode solution
under this background becomes chiral: the undisplayed complex
conjugate, ``c.c.'' corresponding to each displayed matter in
Eq.~(\ref{branching}) appears as just an anti-particle state to
form chiral matter \cite{Berkooz etal,CIM,KatzVafa, Abe etal}.
Second, considering an instanton background in heterotic dual
theory, the actual physical degree is only that modded out by
$S_4$ monodromy. It can be realized by a ``spectral cover.''

\subsection{Monodromy}

To study its consequence, it is convenient to deal with the
weights of $\bf 4$ as $\{t_1,t_2,t_3,t_4\}$. Our $S_4$ is the
permutation group shuffling all of these four weights. We can also
associate U(1)$_X$ charged SU(4)$_\perp$ singlet ${\bf 1}_X$ as
$\{t_5\}$. It is understood as embedding
SU(4)$_\perp\times$U(1)$_X \subset$ SU(5)$_\perp$, under which
${\bf 4} + {\bf 1}_X \to {\bf 5}$.

The U(1)$_X$ quantum numbers subscripted in Eq.~(\ref{branching})
are correctly reproduced by assigning
\begin{eqnarray} \label{X}
X=(1,1,1,1;-4)
\end{eqnarray}
in the $(t_1,t_2,t_3,t_4,t_5)$ basis. Modding out by $S_4$, we have two kinds of ${\bf 10}$ representations
$$ {\bf 10}_i :\{t_1,t_2,t_3,t_4\} ~,\quad {\rm and} \quad ~ {\bf 10}_{-4}: \{t_5\} ~. $$
Likewise, $S_4$ distinguishes two kinds of $\bf 5$'s,
\begin{equation*} \begin{split}
 {\bf \overline 5}_h &~:~\{t_1+t_2,t_1+t_3,t_1+t_4,t_2+t_3,t_2+t_4,t_3+t_4\} ~, \\
 {\bf \overline 5}_i &~:~\{t_1+t_5,t_2+t_5,t_3+t_5,t_4+t_5 \} ~.
  \end{split}
\end{equation*}
In the same way, we can identify the SU(5) singlets. We see that
the matter fields naturally compose the SO(10) multiplets.

The Yukawa couplings of Eq.~(\ref{Yukawa}) are deduced from the
gauge invariant Chern--Simons interactions \cite{BHVI}, having the
structure
\begin{equation} \label{invariance} \begin{split}
{\bf 10}_i{\bf 10}_j{\bf 5}_{h}&~:~ (t_m) + (t_n) + (-t_m - t_n) =0 ~, \\
{\bf 10}_i{\bf\overline{5}}_{j}{\bf\overline{5}}_{h}&~:~
(t_m)+(t_n+t_5)+(t_p+t_q) = 0~, \\
{\bf 1}_{i}{\bf\overline{5}}_{j}{\bf 5}_{h} &~:~
(t_m - t_5) + (t_n+t_5) +(-t_m - t_n) =0 ~,\\
{\bf\overline{10}}_H{\bf\overline{ 10}}_H{\bf 10}_i{\bf 10}_j
&~:~(-t_m)+(-t_n)+(t_m)+(t_n) = 0 ~,
  \end{split}
\end{equation}
where all the indices, $m,n,p,q$ run over $1,2,3,4$ and are
different. Later we will distinguish ${\bf 10}_i$ and ${\bf 10}_H$ only by
the vacuum expectation value (VEV): the ${\bf 10}$ developing a nonzero
VEV is regarded as ${\bf 10}_H$. We cannot distinguish them by introducing another monodromy, since it is simply a vector representation under U(4)$_\bot$.
To justify the second line of Eq.~(\ref{invariance}), we have to
impose the traceless relation,
\begin{equation} \label{unimodular}
 t_1 + t_2 + t_3 + t_4 + t_5 = 0 ~.
\end{equation}
Hence, the `trace part' of U(4) is cancelled by U(1)$_X$. Thus,
sometimes the structure group is suggestively denoted by
S[U(4)$_{\perp}\times $U(1)$_X]$. However, we also find that there
are couplings like
\begin{equation} \label{pdecayinv}
 {\bf 10}_1{\bf 10}_1{\bf 10}_1{\bf\overline{5}}_{-3} ~:~ (t_i)+(t_j)+(t_k)+(t_l+t_5) =
 0~,
\end{equation}
yielding proton decay operators, $d^c d^c u^c$ in
Eq.~(\ref{Rp-viol}) and $q q q l$ in Eq.~(\ref{D5pdecay}) at tree
level.

The best remedy is to further decompose S[U(4)$_\perp \times
$U(1)$_X] \to $S[U(3)$_\perp\times $U(1)$_Z \times $U(1)$_X]$ by singling
out $t_4$, and introduce $S_3$ monodromy on the U(3) part.
Observing the quantum number, it is easy to find the spectrum,
summarized in Table \ref{tb:matter}. There is a new commutant
group U(1)$_Z$, generated by
\begin{eqnarray} \label{Z}
Z=(1,1,1,-3,0)
\end{eqnarray}
in the same basis. Since ${\bf 10}_1$ representation can take only
one of weights $t_1, t_2, t_3$ {\em except} $t_4$ and we assign
${\bf \overline 5}_{-3}$ matter as $\{t_i + t_5, i=1,2,3\}$, it is
impossible to satisfy (\ref{pdecayinv}). If there is no ${\bf
\overline 5}_{-3}':\{t_4+t_5\}$ due to $G$-flux, shown in Table
\ref{tb:matter}, we have no dangerous dimension 4 operators of
Eqs.~(\ref{Rp-viol}) and the dimension 5 operators of
Eq.~(\ref{D5pdecay}).

\begin{table}
\begin{tabular}{c|c|c|c}
\hline Matter & Matter Curve & Homology Class & Net \# of Families \\
\hline ${\bf 10}_{1}$ & $\prod_it_i\rightarrow 0$ &
$\sigma \cap (\eta-3c_1)$ &~ $-(\lambda\eta-\frac13\zeta) \cdot (\eta-3c_1)=3$
\\
${\bf 10}_{1}^\prime$ & $t_4\rightarrow 0$ & $\sigma \cap (-c_1)$ &
$c_1 \cdot \zeta = 0$
\\
${\bf \overline{5}}_{-3}$ & $\prod_i(t_i+t_5)\rightarrow 0$ &
$\sigma \cap (\eta-3c_1)$ &~ $-(\lambda\eta-\frac13\zeta) \cdot (\eta-3c_1)=3$\\
${\bf \overline{5}}_{-3}^\prime$ & $t_4+t_5\rightarrow 0$ &
$\sigma \cap (-c_1)$ & $c_1 \cdot \zeta =0$\\
${\bf 1}_5$ & $\prod_i(t_i-t_5)\rightarrow 0$ &
$\sigma \cap (\eta-3c_1)$ &~ $-(\lambda\eta-\frac13\zeta) \cdot (\eta-3c_1)=3$
\\
${\bf 1}_5^\prime$ & $t_4-t_5\rightarrow 0$ & $\sigma \cap (-c_1)$ &
$c_1 \cdot \zeta = 0$
\\
${\bf 5}_{-2}$ ($\equiv{\bf 5}_{h}$) ~&~
$\prod_{i,j}(-t_i-t_j)\rightarrow 0$ ~&~
$(2\sigma+\eta) \cap (\eta-3c_1)$
~&~ $-(\lambda\eta+\frac23\zeta)\cdot (\eta-3c_1)=1$ \\
${\bf \overline{5}}_{2}^\prime$ ($\equiv{\bf \overline{5}}_{h}$) &
$\prod_i(t_i+t_4)\rightarrow 0$ & $\sigma \cap (\eta-3c_1)$
&~ $-(\lambda\eta+\frac23\zeta)\cdot (\eta-3c_1)=1$
\\
${\bf 1}_0$ & $\prod_i(t_i-t_4)\rightarrow 0$ &
$\sigma \cap (\eta-3c_1)$ &~ $-(\lambda\eta-\frac43\zeta) \cdot (\eta-3c_1)=5$
\\
${\bf 10}_{-4}$ & $t_5\rightarrow 0$ & $\sigma \cap (-c_1)$ & $0$
\\ \hline
\end{tabular}
\caption{ ~Field spectrum in the F-theory model of flipped SU(5).
Fluxes $\lambda(3\sigma_\infty-\eta)+\frac13\zeta$ and $-\zeta$
are turned-on on $C^{(a)}$ and $C^{(b)}$, respectively. We take
$\lambda=\frac16$, $\eta \cdot (\eta-3c_1)=-14$, $\eta \cdot \zeta=2$, and
$c_1 \cdot \zeta=0$ for obtaining three families of matter and only one
pair of the electroweak Higgs. } \label{tb:matter}
\end{table}

\subsection{Matter Curves} \label{s:mcurves}

To have four dimensional $N=1$ SUSY, we compactify F-theory on an
elliptic Calabi--Yau fourfold. Our SU(5)$\times$ U(1)$_X$ gauge
group is located at a codimension 1 complex surface $S_{\rm GUT}$
in the base $B$ of the elliptic fiber. In analogy to perturbative
Type IIB string, we interpret that a stack of sevenbranes wraps
$S_{\rm GUT}$ and the rest of the direction to be our 4 noncompact
spacetime dimensions.

In this subsection, only the structure group will be described,
and the concrete realization of $S_{\rm GUT}$ will be given in the
following subsection. To obtain the transformation property
reflecting monodromy, we introduce a spectral cover \cite{FMW}. It
encodes the symmetry breaking information. The information on the
structure group S[U(3)$\times$ U(1)$_Z\times$ U(1)$_X]$ is contained
in the spectral covers $C^{(a)} \cup C^{(b)} \cup C^{(d)}$. It is
described by the algebraic equation,
\begin{eqnarray}\label{speccover}
 P_a P_b P_c \equiv (a_0U^3+a_1U^2V+a_2UV^2+a_3V^3)(b_0U+b_1V)(d_0U+d_1V)=0
 ~,
\end{eqnarray}
where each factor corresponds to the cover with the same index.
Here we consider a conventional dual space to $B$ via
projectivization:
$$\v Z = {\mathbb P}(K_S \oplus {\cal O}) \stackrel{\pi}{\to} S_{\rm GUT}~,$$
where $K_S$ and $\cal O$ indicate the canonical and trivial
bundles on $S_{\rm GUT}$, respectively. $U$ and $V$ parameterize
respectively the zero section $\sigma$ and the section at infinity
$\sigma_\infty \equiv \sigma+\pi^* c_1(S_{\rm GUT})$ such that
$\sigma \cap \sigma_\infty=0$ (see e.g. Ref. \cite{Marsano etal}).
In other words, $U=0$ is the location of $S_{\rm GUT}$. On $S_{\rm
GUT}$, hence, $a_m$ are sections of $-t +(6-m) c_1$, where $c_1$
and $-t$ symbolize the first Chern classes of the tangent bundle
of $S_{\rm GUT}$ and the normal bundle to $S_{\rm GUT}$ in $B$.
Also both $b_1/b_0$ and $c_1/c_0$ transform as $-c_1$.

We can relate weights ${\bf 3}:\{t_1,t_2,t_3\},\ {\bf
1}_Z:\{t_4\},\ {\bf 1}_X:\{t_5\}$ of the structure group and the
positions of the spectral covers as
\begin{equation} \begin{split}
 a_1/a_0 &\sim t_1 + t_2 + t_3 ~, \\
 a_2/a_0 &\sim t_1 t_2 + t_1 t_3 + t_2 t_3 ~, \\
 a_3/a_0 &\sim t_1 t_2 t_3 ~, \\
 b_1/b_0 &\sim t_4 ~, \\
 d_1/d_0 &\sim t_5 ~,
  \end{split}
\end{equation}
reflecting the $S_3$ monodromy. The unimodular condition
Eq.~(\ref{unimodular}) implies $a_0 b_0 d_1 + a_0 b_1 d_0 + a_1
b_0 d_0 = 0$, with which the three covers can not be independent.
To be consistent with the Green-Schwarz relation in 6 dimensions,
$b_0$ and $d_0$ should be the trivial sections on $S_{\rm GUT}$
\cite{Sadov,DWI,KSC}. So we set
\begin{eqnarray} \label{6dgsrel}
b_0=d_0=1 ~.
\end{eqnarray}
Thus, the traceless condition of SU(5)$_\bot$ becomes
\begin{eqnarray}\label{traceless}
  a_1=-a_0(b_1+d_1) ~.
\end{eqnarray}

The matter field appears at a curve, along which the gauge
symmetry is enhanced \cite{KatzVafa}. As discussed before, the
off-diagonal components from the branching yield chiral matter. In
$\v Z$, a certain factor of the spectral cover (and combinations
thereof) intersect $S_{\rm GUT}$ along such matter curves. From
the weight vectors, as presented in Table \ref{tb:matter}, one can
see which combinations of the spectral covers give the specific
matter fields. For instance, ${\bf 10}_1$ matter field associated
with $t_1 t_2 t_3 \to 0$, is localized at the curve $\{a_3 = 0\}$.
It is obtained from
\begin{equation} \label{10matter}
 C^{(a)} \cap \sigma = \pi_a^*(\eta-3c_1) \cap \sigma ~.
\end{equation}
Setting $U=0$ in the equation for $C^{(a)}$, we indeed obtain the
equation $a_3 = 0$ on $S_{\rm GUT}$. Note that $V$ can not be zero
when $U=0$.

The Higgs field ${\bf  5}_{-2}$ appears as $\prod_{1\le i,j \le 3}
(t_i+t_j) \to 0$. Since both $t_i$ and $t_j$ are inside $C^{(a)}$,
we expect that the corresponding curve comes from the intersection
$C^{(a)} \cap \tau C^{(a)}$. They are the common solutions of
$P_a(V)=0$ and $P_a(-V)=0$, or
$$U(a_0U^2+a_2V^2)=0 ~~\text{ and }~~ V(a_1U^2+a_3V^2)=0~.$$
Since $C^{(a)} \cap \sigma$ or $U=a_3=0$ correspond to ${\bf
10}_1$, which has been already counted, now we don't consider this
possibility for ${\bf  5}_{-2}$. Also another redundant solution
is $V=a_0=0$. We also drop it since $V=0$ is infinitely far from
$S_{\rm GUT}$. Thus, the remaining equation we should solve for
$C^{(a)} \cap \tau C^{(a)}$ is $a_0U^2+a_2V^2=a_1U^2+a_3V^2=0$.
However, still we have an irrelevant solution, $V^2=a_0=0$. Here
it should be noted that $a_1=0$ by Eq.~(\ref{traceless}) if
$a_0=0$. Hence we should drop it also. Then the remaining
solution, which corresponds to ${\bf  5}_{-2}$, becomes associated
with the following homology class;
\begin{eqnarray} \label{fivematter}
{\bf \overline 5}_{2}~:~~(\pi^*\eta+2\sigma)\cap \big\{
\pi^*(\eta-c_1)+2\sigma \big\}-2\sigma_\infty\cap\pi^*\eta =
(2\sigma+\pi^*\eta)\cap\pi^*(\eta-3c_1).
\end{eqnarray}
Here we used $\sigma \cap \sigma = -c_1 \cap \sigma$.


Similarly, by surveying the index structure one can see where the
other matter curves are located. For the curves inside $C^{(i)}
\cap \tau C^{(j)},i,j=a,b,d$, we look for the common solution of
$P_{i}(V)=0$ and $P_{j}(-V)=0$, drop the redundant part, and read
off the homology. The results are
\begin{align}
{\bf 10}_{1}^\prime &~:~ \sigma\cap\pi^*(-c_1) ~\in ~C^{(b)} \cap \tau C^{(b)} ~, \label{tenprime} \\
{\bf 10}_{-4} &~:~ \sigma\cap\pi^*(-c_1) ~\in ~C^{(d)} \cap \tau C^{(d)} ~, \\
{\bf \overline{5}}_{2}^\prime &~:~\sigma\cap\pi^*(\eta-3c_1) ~\in ~C^{(a)}\cap\tau C^{(b)} ~,\\
{\bf\overline{5}}_{-3} &~:~\sigma\cap\pi^*(\eta-3c_1) ~\in ~C^{(a)}\cap\tau C^{(d)} ~, \\
{\bf\overline{5}}_{-3}^\prime &~:~\sigma\cap\pi^*(-c_1) ~\in
~C^{(b)}\cap\tau C^{(d)} ~.
\end{align}
For the curves inside $C^{(i)} \cap C^{(j)}$, we also find the
common solutions of $P_{i}(V)=0$ and $P_{j}(V)=0$, but this is
meaningful only for $i \ne j$. The results are
\begin{align}
{\bf 1}_0 &~:~\sigma\cap\pi^*(\eta-3c_1) ~\in ~C^{(a)}\cap C^{(b)} ~, \\
{\bf 1}_{5}&~:~\sigma\cap\pi^*(\eta-3c_1) ~\in ~C^{(a)}\cap C^{(d)} ~, \\
{\bf 1}_{5}^\prime &~:~\sigma\cap\pi^*(-c_1) ~\in ~C^{(b)}\cap
C^{(d)} ~. \label{onematter}
\end{align}
The number of generations in 4 dimension presented in the last
column in Table \ref{tb:matter} is determined after fluxes are
turned-on. We will discuss it in subsection D.

\subsection{Elliptic Equation}

By definition of F-theory, the Calabi--Yau fourfold contains a
torus, which is described by the elliptic equation,
\begin{equation} \label{ellfiber}
  y^2 + \a_1 xy + \a_3 y = x^3 + \a_2 x^2 + \a_4 x + \a_6 ~.
\end{equation}
The coefficients $\a_m$ are the sections of $(-m)$-th power of the
canonical bundle $K_B$ for the vanishing first Chern class of the
Calabi--Yau fourfold. By reading off the dependence of $\a_m$'s on
the normal direction to $S_{\rm SUT}$, we can identify the gauge
group on $S_{\rm GUT}$. Tate's classification for the simple
groups is tabulated e.g. in Ref. \cite{Bershadsky etal}. For a
semi-simple (possibly plus Abelian) group, we can construct a
similar equation using the information on the spectral cover
\cite{Choi}. Expanding the spectral cover Eq.~(\ref{speccover}),
we have the special combinations of $a_m,b_m$ and $d_m$ as the
coefficients of $U^k V^{5-k}$. These combinations enter as the
coefficients of the elliptic equation Eq.~(\ref{ellfiber}):
\begin{equation} \label{eleqcoeff} \begin{split}
 \a_1 &= -a_3 b_1 d_1 + O(z), \\
 \a_2 & = (a_2 b_1 d_1 + a_3 b_1 + a_3 d_1) z + O(z^2), \\
 \a_3 & = -(a_1 b_1 d_1 + a_2 b_1 + a_2 d_1 + a_3) z^2 + O(z^2), \\
 \a_4 & = (a_0 b_1 d_1 + a_1 b_1 + a_1 d_1 + a_2) z^3 + O(z^4), \\
 \a_6 & = a_0 z^5 + O(z^6),
  \end{split}
\end{equation}
where $z$ parameterizes the normal space to $S_{\rm GUT}$ in $B$.
The other parameters in Eq.~(\ref{eleqcoeff}) are those appearing
in Eq.~(\ref{speccover}), the spectral cover for the
SU(5)$\times$U(1)$_X\times$U(1)$_Z$ group. For the `unfactorized'
SU(5)$\times$U(1)$_X$ case, one can obtain the corresponding
equation in a similar way.

Completing the square in $y$ on the left-hand-side of
Eq.~(\ref{ellfiber}), the discriminant of the remainder in $x$
takes the following form;
\begin{equation} \begin{split}
 \Delta =& a_3^4 b_1^4 d_1^4 (b_1 + d_1) (a_3 + a_2 b_1 + a_2 d_1)(a_3 +a_2 b_1 - a_0 b_1^2 - a_0 b_1^2 d_1)(a_3 + a_2 d_1 - a_0 b_1 d_1^2) z^5 \\
 &+ a_3^2 b_1^2 d_1^2 P z^6 + Q z^7 + O(z^8) ~,
  \end{split}
\end{equation}
where we used the traceless constraint Eq.~(\ref{traceless}) to
eliminate $a_1$ and the coefficients $P$ and $Q$ are not
proportional to $a_3,b_1$ and $d_1$. The coefficient of $z^5$ is
factorized to give various matter curve equations on $S_{\rm
GUT}$, obtained in section \ref{s:mcurves}. One obvious limit is
$d_1 \to 0$, in which the gauge symmetry is enhanced to $O(z^7)$,
which yields SO(10). Other limits such as $a_3 \to 0$ or $b_1 \to
0$ gives also SO(10) enhancements, but they are not along the
chain of $E_n$ series unifications. The specially tuned form of
Eq.~(\ref{eleqcoeff}) indicates a larger gauge symmetry than
generic SU(5), which must be SU(5)$\times$U(1)$_X$. We will
analyze further this symmetry later.

\subsection{Fluxes and Chiral Spectrum}

The matter curves obtained in the previous section span 6
dimensional world volumes. To obtain 4 dimensional chiral
spectrum, we turn-on $G$-flux \cite{FMW}. Since the GUT group
SU(5)$\times$U(1)$_X$ is broken by the Higgs scalar, we only need
to turn-on fluxes on the spectral cover, not on the GUT
sevenbranes. To keep $\sin^2\theta_W^0=\frac38$, we should
preserve the SO(10) unification relation. Its commutant group
under $E_8$ is SU(4)$_\perp$, and the SU(3)$_\perp$ and U(1)$_Z$
covers are identified as $C^{(a)}$ and $C^{(b)}$, respectively.
Hence, we turn-on the universal fluxes only on $C^{(a)}$ and
$C^{(b)}$ to preserve the SO(10) structure.

First, we turn-on a line bundle $\cal N$ on $C^{(a)}$, inducing
the U(3) vector bundle $V=\pi_{a*} {\cal N}$ on $S_{\rm GUT}$:
\begin{equation} \textstyle
 \Gamma_a = \lambda \left\{3\sigma - \pi_a^* (\eta - 3c_1 )\right\}
 + \frac13  \pi_a^* \zeta
\end{equation}
with the projection $\pi_a:C^{(a)} \to S_{\rm GUT}$. The trace
part is $\zeta = c_1(V)$, and it is cancelled by a line bundle on
$C^{(b)}$ \cite{Blumenhagen etal},
\begin{equation}
 \Gamma_b = - \pi_b^* \zeta ~
\end{equation}
with the projection $\pi_b:C^{(b)} \to S_{\rm GUT}$.
%

We have the quantization condition for $\cal N$ \cite{FMW},
\begin{equation} \label{quantize}
 c_1 ({\cal N}) = \frac12 \left\{ -c_1(C^{(a)})+\pi_a^* c_1\right\}
 + \Gamma_a ~\in ~H^2( C^{(a)},\Z)~.
\end{equation}
From the adjunction formula for $\v Z$, we have
\begin{equation}
 -c_1(C^{(a)}) + \pi_a^* c_1 =\sigma + \pi_a^* \left(\eta -
 c_1\right) ~.
\end{equation}
Thus, the quantization condition for $\cal N$,
Eq.~(\ref{quantize}) provides the following nontrivial
constraints;
\begin{equation} \label{fluxquantization} \textstyle
 3(\frac12+\lambda) \in \Z, \quad -(\lambda -\frac12) \eta
 + (3\lambda - \frac12)c_1 +\frac13 \zeta
 ~\in ~H^2(S_{\rm GUT},\Z) ~.
\end{equation}
The $C^{(d)}$ cover responsible for U(1)$_X$ is a single cover,
and so we can turn-off the flux. Then the unwanted ${\bf 10}_{-4}$
becomes vector-like, and so it can be removed from the low energy
field spectrum.
From now on, we will drop the symbol of pullback `$\pi^*$' for
simplicity, unless they are unclear.

The net numbers of the chiral fields are calculated using
Riemann-Roch-Hirzebruch index theorem \cite{Donagi:2004ia,Blumenhagen:2006wj}
\begin{equation}
n(R) \equiv  n_R - n_{\overline R} = \Sigma_R \cap \Gamma |_\sigma
\end{equation}
where $\Sigma_R$ is the matter curve inside $\check Z$, shown in (\ref{10matter}),(\ref{fivematter}) and (\ref{tenprime})-(\ref{onematter}).
Specifically we have
\begin{eqnarray}
&&n({\bf 10}_{1}) =n({\bf \overline{5}}_{-3}) = n({\bf 1}_{5})
\nonumber \\
&&\quad\quad\quad~=\left[\left(\sigma\cap(\eta-3c_1)\right)\cap\left(\lambda
(3\sigma_\infty-\eta)+\frac13\zeta\right)
+\left(\sigma\cap(-c_1)\right)\cap(-\zeta)\right]_\sigma
\nonumber
\\
&&\quad\quad\quad~
=-\left(\lambda\eta-\frac13\zeta\right) \cdot \left(\eta-3c_1\right) +c_1 \cdot \zeta~, \\
&&n({\bf \overline{5}}_{2})=
\left[\left(\left(2\sigma+\eta\right)\cap\left(\eta-3c_1\right)\right)\cap
\left(\lambda\left(3\sigma_\infty-\eta\right)+\frac13\zeta\right)
\right]_\sigma \nonumber \\
&&\quad\quad~=\left(\lambda\eta+\frac23\zeta\right) \cdot (\eta-3c_1)~
 ,\\
&&n({\bf
\overline{5}}_{2}^\prime)=\left[\left(\sigma\cap(\eta-3c_1)\right)\cap
\left(\eta(3\sigma_\infty-\eta)+
\frac13\zeta-\zeta\right)\right]_\sigma
\nonumber \\
&&\quad\quad~ =-\left(\lambda\eta+\frac23\zeta\right) \cdot (\eta-3c_1)
~,\\
&&n({\bf 1 }_0)=\left[\left(\sigma\cap(\eta-3c_1)\right)\cap
\left(\eta(3\sigma_\infty-\eta)+
\frac13\zeta+\zeta\right)\right]_\sigma
\nonumber \\
&&\quad\quad~ =-\left(\lambda\eta-\frac43\zeta\right) \cdot (\eta-3c_1)
~.
\end{eqnarray}
Here the intersection is done in the $\check Z$ space, and the dot product is done on $S_{\rm GUT}$.
Thus, the existence of three families of the SM matter and only
one pair of the vector-like electroweak Higgs fields require
$-\left(\lambda\eta-\frac13\zeta\right) \cdot \left(\eta-3c_1\right)
+c_1 \cdot \zeta=3$ and
$-\left(\lambda\eta+\frac23\zeta\right) \cdot (\eta-3c_1)=1$. To kill the
unwanted superpotential terms, ${\bf 10}_1^{(\prime)}{\bf
10}_1^{(\prime)}{\bf
10}_1^{(\prime)}{\bf\overline{5}}_{-3}^{(\prime)}$ and ${\bf
10}_H{\bf 10}_1^{(\prime)}{\bf
10}_1^{(\prime)}{\bf\overline{5}}_{-3}^{(\prime)}$, as mentioned
above, the {\it matter} fields associated with $t_4\rightarrow 0$
i.e. ${\bf 10}_1^\prime$, ${\bf\overline{5}}_{-3}^\prime$, and
${\bf 1}_5^\prime$ should be absent at low energies. Hence, we
take
\begin{eqnarray} \label{3Fcondi}
\lambda\eta \cdot (\eta-3c_1)=-\frac{7}{3}~,~~\eta \cdot \zeta =2~,~~{\rm
and}~~c_1 \cdot \zeta=0
\end{eqnarray}
Moreover, the absence of a flux on $C^{(d)}$ leaves the exotic
field ${\bf 10}_{-4}$ vector-like. The zero modes of the chiral
field spectrum are summarized in Table~\ref{tb:matter}.

For constructing a a local
model, a necessary condition is that the four cycle $S_{\rm GUT}$
is a del Pezzo surface $dP_n$ \cite{BHVI}: $S_{\rm GUT}$ should be shrinkable inside the ambient
space. In global model it is not necessary but del Pezzo surface is easy to realize as a projective variety. The first constraint in Eq.~(\ref{fluxquantization}) is
easily fulfilled by taking $\lambda=\frac16$. Then the second
constraint in Eq.~(\ref{fluxquantization}) implies
\begin{eqnarray}
%
\frac13\left(\eta+\zeta\right) \in H_2 (S_{\rm GUT},\Z)
\end{eqnarray}
for $\lambda=\frac16$. We can find $\eta$ and $\zeta$ satisfying
Eqs.~(\ref{3Fcondi}), e.g. just if $S_{\rm GUT}=dP_2$, namely, the
canonical class is given by $-K_S=c_1=3H-E_1-E_2$:
\begin{eqnarray}
\eta = 2H ~,~~\zeta = H-3E_1 ~,
\end{eqnarray}
where $H$ and $E_i$ ($i=1,2$) denote the hyperplane divisor and
exceptional divisors, respectively, satisfying
\begin{eqnarray}
H\cdot H=1~,~~E_i\cdot E_j=-\delta_{ij}~,~~{\rm and}~~H\cdot E_i=0
~.
\end{eqnarray}
The global embedding is easily done by borrowing the $dP_2$ construction in \cite{Marsano:2009ym}.

\section{Abelian Symmetry}

\subsection{U(1)$_X$ as Gauge Symmetry}

There is an issue concerning U(1) gauge group
\cite{Hayashi:2010zp,Grimm:2010ez,Marsano:2010ix}. Since there is
only one Cartan subalgebra, we cannot identify it geometrically.
The Cartan subalgebra are obtained by reducing the three-form
field along two-cycles, and their field strength satisfies
\begin{equation}
 G = \sum F \wedge \omega, \quad \omega \in H^2(\rm{CY_4},\Z),
\end{equation}
using collective notation, where $G$ is four-form field strength
analogous to one in M-theory and two forms $\omega$ are not in the
three-base or the elliptic fiber. If we turn-on a line bundle on
this cover, then it potentially makes the corresponding gauge
boson massive by the St\"uckelberg mechanism. From the interaction
involving $G$ we have the induced action \cite{DWII},
\begin{equation} \label{stuckelberg}
 \int_{{\mathbb R}^{1,3}} F_X \wedge c_2^{(i)}
 \tr X^2 \int_S c_1(L_X) \wedge \iota^* \omega_i ~.
\end{equation}
Note that the contribution from U(1)$_X$ charges is proportional
to $\tr X^2$. Even if there is no 4 dimensional gauge and
gravitational anomalies proportional to $\tr X$ or $\tr X^3$,
still there is a room for massive gauge bosons. In our situation we do not turn on flux along the U(1)$_X$,
$L_X =0$, there the gauge boson is massless from (\ref{stuckelberg}).

We can check that there is no U(1)$_X$ gauge and gravitational
anomalies
$$
 \sum_R n(R) \dim (R) X^3=0, \quad \sum_R n(R)  \ell(R) X=0, \quad \sum_R  n(R) \dim (R) X  = 0,
$$
where $n(R)$ is the net number of chiral minus antichiral
generations for $R$. $\ell(R)$ denotes the Dynkin index defined as
\begin{equation}\label{dynkin}
 \tr_R T^a T^b = \ell(R) \delta^{ab} ~,
\end{equation}
for the generators $T^a$ of a Lie group $G$. Since there is no
missing charged matter, it seems that this U(1)$_X$ can be fully
understood in local description. A supporting argument is, in the
case where U(1)$_X$ is protected by a larger nonabelian gauge
group (e.g. as in Ref. \cite{Choi}) e.g. SO(10) as in our case, we
can arbitrarily shrink the two-cycle where U(1)$_X$ is local
enough, being caught around $S_{\rm GUT}$. There is no new
monodromy mixed with the cycles outside $E_8$. In some
literatures, the condition for six dimensional anomaly cancellation
conditions Eq.~(\ref{6dgsrel}) were not met, and so the
identification of SU(5) singlets carrying U(1)$_X$ charges was
failed.

\subsection{U(1) Normalization}

We explain how the normalization of U(1)$_X$ is determined when
flipped SU(5) is embedded in SO(10). Formerly, our definition of
U(1)$_X$ in Eq.~(\ref{X}) does not rely on the SO(10), but
embedded in the SU(5)$_\perp$, which is the commutant group of the
GUT SU(5) in $E_8$. We will see how they are related.

In dealing with normalization, the Dynkin index defined in
Eq.~(\ref{dynkin}) is useful. Once we fix $\ell(R)$ for one kind
of representation $R$, it fixes the normalization of all the
generators of the group $G$. For example, the complex conjugate
representation $\overline R$ of $R$ has the relation
$\ell(\overline R)=\ell(R)$.

Considering a subgroup $H$ of $G$ and the commutant group
$\Gamma$, there is a property
\begin{equation} \label{indexprop}
 R \to \sum (R_{H}, R_{\Gamma}), \quad \ell(R) = \sum \ell(R_H) \dim(R_{\Gamma}) ~.
\end{equation}
As an example, consider SU($n$) and its subgroup SU(2). Fixing
$\ell({\bf 2})=\frac12$ for SU(2), also fixes $\ell({\bf
n})=\frac12$ for the fundamental representation of SU($n$). It is
easily shown by relation Eq.~({\ref{indexprop}) and using the fact
that the singlet is neutral under the group $\ell({\bf 1})=0$. The
relation Eq.~(\ref{indexprop}) is unique so that the converse also
holds. Starting from any group $G$, we can show the same relation
to its SU(2) subgroup in {\em any direction}. An important
consequence is that
\begin{equation} \label{SUnNorm}
\text{
 the generators of any SU-type subgroup of $G$ have the same normalization}
\end{equation}
in the fundamental representation.

Now, consider SU(5)$\times$U(1)$_X$ subgroup of SO(10). This
embedding is easily understood by conventional `complexification.'
Since ${\bf }$$\{{\bf 5}_{-2},{\bf\overline{5}}_{2}\}$ in SU(5)
are embedded in a vector representation $\bf 10$ of SO(10), the
{\em same} U(1)$_X$ generator has two {\em different
representations}, $T_X^{\bf 10}$ with respect to ${\bf 10}$ of
SO(10) and $T_X^{\bf 5}$ to ${\bf 5}$ of SU(5), for example,
\begin{equation} 
  T_X^{\bf 10} =   \begin{pmatrix} 0 & -1 \\ 1 & 0 \end{pmatrix} \otimes iT_X^{\bf 5 \prime}
  \stackrel{\text{diagonalization}}{\longrightarrow} \begin{pmatrix} 1 & 0 \\ 0 & -1 \end{pmatrix}\otimes  T_X^{\bf 5} .
\end{equation}
Explicitly, $T_X^{\bf 5}=  \rm diag(-2,-2,-2,-2,-2)$. This shows,
once we fix the normalization $\ell({\bf \overline 5})=\frac12$ of
SU(5), SO(10) vector has the normalization $\ell({\bf 10})=1$, as
it should be from Eq.~(\ref{indexprop}). In other words,
\begin{equation} \label{su5so10}
 |(-2,-2,-2,-2,-2;2,2,2,2,2)|^2/\ell({\bf 10})=|(-2,-2,-2,-2,-2)  |^2/\ell({\bf 5}) = 40,
\end{equation}
in the SU(5) basis. It implies the normalized generator is $
\frac{1}{\sqrt{40}} T_{X}^{ \bf 5}$.

In our case, U(1)$_X$ of flipped SU(5) is embedded in the
SU(5)$_\perp$, which is the commutant group of the GUT SU(5) in
$E_8$, as shown in Eq.~(\ref{X}). So at first sight its generator
$X$ seems not be related to the previous SO(10). However U(1)$_X$
is the common intersection between SO(10) and SU(5)$_\perp$, so we
observe that $X$ and $T_X^{\bf 5}$ are the {\em same} generators
with merely different representations. This is of course
understood as being $E_8$ generator. Here we check explicitly
their normalizations are the same. A particular case of
Eq.~(\ref{SUnNorm}) is that, fixing $\ell({\bf \overline
5})=\frac12$ for the SU(5), it should be that $\ell({\bf \overline
5}_\perp)=\frac12$ for the other SU(5)$_\perp$. As a generator of
SU(5)$_{\perp}$, $X$ should be replaced with the one with
normalization $\tr \tilde X^2 = \frac12$. Indeed,
\begin{equation} \label{Xnorm}
 |(-2,-2,-2,-2,-2)|^2/\ell({\bf 5})=|(1,1,1,1,-4)|^2/\ell({\bf 5}_\perp)=40.
\end{equation}
Since we independently identified the generators,
Eq.~(\ref{Xnorm}) gives a nontrivial check that
\begin{equation}
 \frac{1}{\sqrt{40}} T_X^{\bf 5}, \quad \frac{1}{\sqrt{40}} X.
\end{equation}
should be different representations of a {\em single} generator of
$E_8$.\footnote { It also follows that the normalized one for
Eq.~(\ref{Z}) should be $\tr \tilde Z^2 = \frac12$ from the
embedding SU(3)$\times$U(1)$_Z \subset$SU(4). Its commutant in
$E_8$ is SO(10), so U(1)$_Z$ is the common intersection of the
SO(10) and the SU(4). Fixing the normalization $\ell({\bf
4})=\frac12$ means also fixing $\ell({\bf 27})=3$. It is done, for
example, by considering the chain $\ell({\bf 27})=\ell({\bf
16}_{1})+\ell({\bf 10}_{-2})+\ell({\bf 1}_4)=2+1+0$. Therefore we
fix the normalization of U(1)$_Z$ generator inside $E_6$, with
respect to the minimal representation $\bf 27$,
$$ |(\underbrace{1,1,\dots,1}_{16},\underbrace{-2,-2,\dots,-2}_{10},4)|^2/\ell({\bf 27})= |(1,1,1,-3)|^2/\ell({\bf 4})=24.
$$
Here the bracing numbers indicate the number of repeated
entries. Also, identifying the SM group as $E_3 \times$U(1)$_Y$,
the commutant to SU(5)$\times$U(1)$_Y$ in $E_8$, from the
embedding to SU(6), we have a similar relation
$$
|(-2,-2,-2,3,3)|^2/\ell({\bf 5}) = |(1,1,1,1,1,-5)|^2/\ell({\bf
6}) = 60~.
$$
}

When flipped SU(5) is broken to the SM gauge group, the gauge
coupling of U(1)$_Y$, $g_Y$ in the SM becomes related to $g_5$ and
$g_X$ of flipped SU(5). We recollect the gauge kinetic terms for
SU(5)$\times$U(1)$_X$ from that of SO(10).
Using the relation $Y=\frac15(T_5+X)$ as in
Eq.~(\ref{hyperY}) and the normalization in Eq.~(\ref{Xnorm}), we
extract the coupling relation,
$$
 - \frac{1}{4g_{SO(10)}^2} \tr_{\bf 10} F^2 \to - \frac{1}{2g_{5}^2} \cdot \frac{1}{5^2} \tr_{\bf 5 } \left( F^2_{T_5} + F^2_{X} \right) =- \frac{1}{4 g_Y^2} F_Y^2,
$$
so that
\begin{equation} \label{g_Y}
\frac{1}{g_Y^2}=\frac{1}{5^2}\cdot\frac{5}{3}\cdot
\frac{1}{g_5^2}+\frac{1}{5^2}\cdot 40 \cdot \frac{1}{g_{X}^2} ~,
\end{equation}
from which we understand $g_X=g_5$ at the GUT scale. We have $g_2=g_5$, since
SU(2)$_L$ in the SM gauge group comes purely from the SU(5) part
of flipped SU(5), thus the bare weak mixing angle at the GUT scale
is
\begin{equation} \label{mixingangle}
{\rm sin}^2\theta_W^0 \equiv
\frac{g_Y^2}{g_2^2+g_Y^2}=\frac{1}{\frac{g_5^2}{g_Y^2}+1}=\frac{3}{8}.
\end{equation}

Why does flipped SU(5) yield the same relation as SU(5)$_{\rm
GG}$?  If $U(1)_X$ is embedded in a simple group of SO(10), there
is a single coupling. The main difference between GG SU(5) and
flipped SU(5) comes from the definition of hypercharges
$$
\textstyle Y_{\rm GG} = \sqrt{\frac35}~ {\rm
diag}(\frac13,\frac13,\frac13,\frac{-1}{2},\frac{-1}{2}), \quad
Y_{\rm F-SU(5)} = \sqrt{\frac35}~ {\rm
diag}(\frac{1}3,\frac{1}3,\frac{1}3,\frac{1}{2},\frac{1}{2}),
$$
in the fundamental representation. They just differ by some signs
of hypercharges while SU(3)$_C\times$SU(2)$_L$ direction intact.
So, from the relation Eq.~(\ref{mixingangle}), the same $g_Y$
normalization gives the same weak mixing angle. The new
hypercharge combination is possible because flipped SU(5) has an
extra component of the U(1)$_X$, as in (\ref{X}). It turns out
that this is the only possible new combination inside
SO(10).

The coupling unification and the weak mixing angle relation is the
same if the symmetry breaking mechanism down to MSSM does not
change gauge coupling. It includes scalar Higgses and Wilson
lines. Such mechanism is sometimes associated with the symmetry of
internal manifold, so heterotic string compactification with a
background gauge bundle, or F-theory with a spectral cover with
the {\em hypercharge untouched} gives the same unification and
weak mixing angle. In F-theory, there is another source of gauge
symmetry breaking: $G$-flux along hypercharge direction gives a
correction to gauge kinetic function thus gauge coupling. So the
relation changes slightly \cite{DWI,distort,ChoiKim}. The
simplest GUT that does not require hypercharge flux is flipped
SU(5), so still the relation is preserved.

\section{Low Energy Effective Theory}

In our F-theory model we obtain 3 {\it net} families of ${\bf
10}_{1}$s, 5 of ${\bf 1}_0$s, and one net pair of $\{{\bf
5}_{-2},{\bf\overline{5}}_{2}\}$. In particular, ${\bf 10}_1$ and ${\bf \overline {10}}_{-1}$ matter representations belong to cohomology $H^0(\Sigma_{\bf 10},K^{1/2}_{\Sigma_{\bf 10}} \otimes V)$ and $H^0(\Sigma_{\bf \overline { 10}},K^{1/2}_{\Sigma_{\bf \overline{10}}} \otimes V^*)$, respectively. At present it is not possible to calculate the individual Euler numbers for them, we suppose $4\times {\bf 10}_1$
and $1\times {\bf\overline{10}}_{-1}$, and regard
${\bf\overline{10}}_{-1}$ and one of ${\bf 10}_1$ as
${\bf\overline{10}}_{H}$ and ${\bf 10}_H$ breaking flipped SU(5),
respectively. On the other hand, we assume that the absolute
number of $\{{\bf 5}_{-2},{\bf\overline{5}}_{2}\}$ is 1.

The VEV distinguishes ${\bf 10}_H$ from ${\bf 10}_i$. Since ${\bf
10}_H$ and ${\bf\overline{10}}_{H}$ have the exactly opposite
gauge quantum numbers, the superpotential admits the following
terms;
\begin{eqnarray}
W\supset M_G{\bf 10}_H{\bf\overline{10}}_H+\frac{1}{M_G}\left({\bf
10}_H{\bf\overline{10}}_H\right)^2+\cdots ~,
\end{eqnarray}
where we assume that the fundamental scale is of the GUT scale as
in the heterotic M theory. From these terms in the superpotential
and and the D-term potential, ${\bf 10}_H$ and
${\bf\overline{10}}_H$ can develop a VEV at a {\it SUSY vacuum},
$\langle{\bf 10}_H\rangle=\langle{\bf\overline{10}}_H\rangle\sim
{\cal O}(M_G)$, satisfying $\partial W/\partial{\bf 10}_H=\partial
W/\partial{\bf\overline{10}}_H=0$ and $\langle{\bf 10
}_H\rangle=\langle{\bf\overline{10}}_H\rangle^*$.
Just based on the symmetries discussed above, the expected low
enrgy superpotential (upto dimension 5) is given by
\begin{eqnarray} \label{bareW}
&&W_{\rm eff}=\left({\bf 10}_H{\bf 10}_H{\bf 5}_h + {\bf 10}_H{\bf
10}_i{\bf 5}_h\right) +{\bf 10}_i{\bf 10}_j{\bf 5}_h
+\frac{1}{M_G}{\bf\overline{10}}_H{\bf\overline{
10}}_H{\bf\overline{5}}_hS^\prime  \\
&&+\left({\bf 10}_H{\bf\overline{5}}_i{\bf\overline{5}}_h+S{\bf
5}_h{\bf\overline{5}}_h\right)+{\bf
10}_i{\bf\overline{5}}_j{\bf\overline{5}}_h + {\bf
1}_i{\bf\overline{5}}_j{\bf 5}_h
+\frac{1}{M_G}{\bf\overline{10}}_H{\bf\overline{ 10}}_H{\bf
10}_i{\bf 10}_j ~, \nonumber
\end{eqnarray}
where we drop the dimensionless coupling constants for simplicity.
As explained earlier, the unwanted terms Eqs.~(\ref{Rp-viol}),
(\ref{D5pdecay}), and (\ref{unwanted}) are absent. $S$ and
$S^\prime$ denote the different linear combinations of five ${\bf
1}_0$s associated with the matter curve of
$\prod_i(t_1-t_4)\rightarrow 0$.

%
%
%

Since ${\bf 10}_i$ and ${\bf 10}_H$ have the same charges in this
model, both ${\bf 10}_H{\bf 10}_H{\bf 5}_h+{\bf 10}_H{\bf
10}_i{\bf 5}_h$ are allowed as seen in Eq.~(\ref{bareW}). It gives
\begin{eqnarray}
\langle\nu^c_H \rangle\left(d^c_H+d^c_1+d^c_2+d^c_3\right)D ~.
\end{eqnarray}
Hence, the mode $\left(d^c_H+d^c_1+d^c_2+d^c_3\right)$ and also
$D$ become heavy, whereas the other 3 modes orthogonal to
$\left(d^c_H+d^c_1+d^c_2+d^c_3\right)$ can be regarded as the
physical d-type quarks. This mixing could suppress the d-type
quark's Yukawa couplings in ${\bf 10}_i{\bf 10}_j{\bf 5}_h$. The
mixing between $d^c_H$ and $d^c_i$ might be helpful for explaining
$m_b/m_t \sim {\cal O}(10^{-2})$.

Similarly, ${\bf 10}_H{\bf\overline{5}}_i{\bf\overline{5}}_h$ and
$S{\bf 5}_h{\bf\overline{5}}_h$ make $l_i$ and $h_d$ mixed:
\begin{eqnarray}
\left\{\langle\nu^c_H \rangle(l_1+l_2+l_3)+Sh_d\right\}h_u = \mu
h_d^\prime h_u ~,
\end{eqnarray}
where $h_d^\prime$ defines the physical d-type Higgs, and $\mu$ is
given by
\begin{eqnarray} \label{mu}
\mu=\sqrt{\langle\nu^c_H \rangle^2 +\langle S\rangle^2} ~.
\end{eqnarray}
Note that $\nu^c_H$ and $S$ are complex fields. Due to the mixing
between $l_i$ and $h_d$, some R-parity violating terms
$q_il_jd^c_j$ and $l_il_je^c_k$ in the superpotential are induced,
but the other term $u^c_id^c_jd^c_k$ is not. $q_il_jd^c_j$ and
$l_il_je^c_k$ violate lepton numbers, but still preserve the
baryon number. Since the dimension 4 proton decay processes are
associated with both $q_il_jd^c_j$ and $u^c_id^c_jd^c_k$, still
the proton can be stable enough.

The smallest one among the upper bounds for the dimension less
lepton number violating couplings is around $10^{-6}$
\cite{Rviol}. It is a similar size of the (R-parity preserving)
electron's Yukawa couplings. In this model, the accidental global
symmetry found at low energies is $Z_3$, under which the MSSM
superfields carry the charges; $q(0), u^c(-1), d^c(1), l(-1),
\nu^c(0), e^c(2), h_u(1), h_d(-1)$ \cite{IbanezRoss}.

The R-parity violating terms arise since both ${\bf 10}_i$ and the
Higgs ${\bf 10}_H$ are parameterized by a single component $t_i$.
Conventionally they are distinguished by imposing $R$-parity in
field theoretic GUT: $(-)$ for the matter fields and $(+)$ for the
Higgs fields. However, there is no known way to embed it to a continuous  symmetry [See also \cite{KuMa}]. Maybe we keep it as an accidental symmetry up to a certain order of perturbation that is suppressed enough \cite{acc}.
In our type factorization, there appears another representation ${\bf
10}_1'$, with which one may attempt to identify as the Higgs. However this leads to proton decay operators
Eq.~(\ref{Rp-viol}). Hence, one may consider a further factorizing
the spectral cover type. However, we will not
pursue this possibility for simplicity of the model, just assuming
the relatively small lepton number violating couplings.

As noticed above, $S$ is composed of the five ${\bf 1}_0$s, and
the VEV of $S$, i.e. all of ${\bf 1}_0$s could remain undetermined
down to low energies. In this case, $\langle S\rangle$ (as well as
$\langle h_u\rangle$ and $\langle h_d\rangle$) would be eventually
fixed by including TeV scale SUSY breaking ``soft terms'' in the
Lagrangian such that $\mu$ of Eq.~(\ref{mu}) becomes of TeV scale.
It implies that $\langle S^\prime\rangle$ in Eq.~(\ref{bareW})
should be of order $\langle{\bf 10}_H\rangle$ ($\sim M_G$),
because $S$ and $S^\prime$ are the different linear combinations
of five ${\bf 1}_0$s. $\langle S^\prime\rangle$ of order $M_G$
induces the second term of Eq.~(\ref{D/Tsplit}).

In this mechanism, a modulus, which is given by a linear
combination of ${\bf 1}_0$s, plays an essential role. However, it
would give rise to the cosmological ``moduli problem.'' Since its
VEV is around the GUT scale but its mass is just of TeV scale, its
decay is not efficient and its oscillation around the true minimum
of the scalar potential is hard to be terminated. This problem
could be resolved by considering the second inflation around TeV
scale temperature (``thermal inflation'') \cite{Lyth}. In this
paper, we don't discuss this issue in details.

\section{Conclusions}

In this paper, we have pointed out that the normalization of the
U(1)$_X$ charges in flipped SU(5) models based on F-theory can be
determined such that ${\rm sin}^2\theta_W^0=\frac38$. It is
because U(1)$_X$ is embedded in a simple structure group
SU(5)$_\perp\subset E_8$. To avoid the dimension 4 and 5 proton
decay, the structure group should split to SU(3)$_\perp$ or
smaller group factors, and extra heavy vector-like pairs $\{{\bf
5}_{-2},{\bf\overline{5}}_2\}$ should be absent. We have proposed
a simple but phenomenologically viable flipped SU(5) model based
on F-theory.


\acknowledgments{}

We are grateful to Eung-Jin Chun, Hirotaka Hayashi, Jonathan Heckman, Jihn E. Kim and
Timo Weigand for discussion.
This research is supported by Basic Science Research Program
through the National Research Foundation of Korea (NRF) funded by
the Ministry of Education, Science and Technology (Grant No.
2010-0009021).



\end{document}